\documentclass[amsmath,floatfix,twocolumn,superscriptaddress,prx,longbibliography]{revtex4-1}
\usepackage{amssymb,amsmath}
\usepackage{graphicx,array}
\usepackage{dcolumn}
\usepackage{bm}
\usepackage{float}
\usepackage{caption}
\usepackage{setspace}
\usepackage{xkeyval}
\usepackage{rotating}

\begin{document}

\title{Atomistic simulation of barocaloric effects}

\author{Claudio Cazorla}
\email{claudio.cazorla@upc.edu}
\affiliation{Departament de F\'isica, Universitat Polit\`ecnica de Catalunya, Campus Nord B4-B5, Barcelona 08034, Spain}

\begin{abstract}
Due to critical environmental issues there is a pressing need to switch from current refrigeration 
methods based on compression of greenhouse gases to novel solid-state cooling technologies. Solid-state 
cooling capitalizes on the thermal response of materials to external fields named ``caloric effect''. 
The barocaloric (BC) effect driven by hydrostatic pressure is particularly promising from a technological 
point of view since typically presents larger cooling potential than other caloric variants (e.g., 
magnetocaloric and electrocaloric effects driven by magnetic and electric fields, respectively). Atomistic 
simulation of BC effects represents an efficient and physically insightful strategy for advancing solid-state 
cooling by complementing, and in some cases even guiding, experiments. Atomistic simulation of BC effects 
involves approaches ranging from computationally inexpensive force fields to computationally very demanding, 
but quantitatively accurate, first-principles methods. Here, we survey several methods and strategies 
involved in atomistic simulation of BC effects like the quasi-harmonic approximation and direct/quasi-direct 
estimation approaches. The Review finalizes with a collection of case studies in which some of these methods 
were employed to simulate and predict original BC effects.
\end{abstract}

\maketitle


\section{Introduction}
\label{sec:intro}
When the isothermal entropy changes associated with fluctuations in the electric, magnetic and structural 
properties of materials are large, $|\Delta S_{T}| \sim 10$--$100$~kJK$^{-1}$kg$^{-1}$, these may render sizable 
caloric effects (e.g., adiabatic temperature changes of $|\Delta T_{S}| \sim 1$--$10$~K). Solid-state cooling is 
an environmentally friendly, highly energy-efficient, and highly scalable technology that exploits caloric 
effects in materials for refrigeration purposes [\onlinecite{cazorla19b,moya14,cazorla17b,cazorla16,scott11,
pecharsky97,bonnot08,shirsath20}]. By applying external fields on caloric materials it is possible to achieve 
reversible temperature shifts that can be integrated in cooling cycles and do not involve the use of greenhouse 
gases. Materials undergoing abrupt structural phase transitions under small hydrostatic pressure are specially 
well suited for solid-state cooling applications based on the barocaloric (BC) effect [\onlinecite{cazorla19b,
moya14,cazorla17b}].  
 
The magnitude of BC effects can be estimated accurately with theoretical and atomistic simulation 
methods. Computer simulations can be used to rationalize the origins of experimentally observed BC 
phenomena since the materials and pressure-induced phase transitions of interest can be accessed at 
the atomic scale under controlled conditions. Moreover, from a resources point of view computer 
simulations are inexpensive. For instance, by using open-source software and modest computational 
resources it is possible, in some cases, to simulate complex BC effects and assess the magnitude of 
the accompanying isothermal entropy and adiabatic temperature shifts. Thus, modeling of BC effects 
can be done systematically in order to complement, and in some cases also guide, the experiments.   

The reliability of computer simulations, however, depends strongly on the simplifications made on the 
adopted structural and interatomic interaction models. Typically, increasing the reliability of the 
structural models comes at the expense of reducing the accuracy in the description of the interatomic 
forces (due to practical computational limitations). For instance, if the simulation approach 
to be employed is accurate first-principles methods most calculations are likely to be performed at 
zero temperature by considering perfectly ordered atomic structures. Such simulation conditions 
obviously differ from the actual experimental conditions. On the other hand, to simulate BC effects
directly at finite temperatures for realistic systems containing crystalline defects and/or other 
inhomogeneities one should use computationally inexpensive interatomic modeling techniques, which may 
suffer from transferability issues and in general have modest predictive power. Fortunately, there are 
well-established simulation strategies that allow to achieve a suitable balance between computational 
accuracy and model reliability and can be used to obtain physically meaningful results. 

In this Review, we survey general computational techniques based on atomistic simulation methods that 
can be used to theoretically estimate and predict BC effects. We start by briefly describing genuine 
first-principles methods, like density functional theory, and other practical approaches that allow
to describe the interactions between atoms in materials, namely, classical interatomic and machine 
learning potentials. Next, several computational techniques that allow to describe phase transitions 
and estimate thermodynamic properties of materials (e.g., free energies and entropies) are summarized 
(e.g., the quasi-harmonic approximation and thermodynamic integration methods). Subsequently, different 
simulation strategies that can be employed for indirect, quasi-direct and direct evaluation of BC effects 
are succinctly explained. We finalize the Review by providing some representative examples in which 
atomistic simulation methods have been used to predict original BC effects in families of materials 
that are technologically relevant, namely, fast-ion conductors [\onlinecite{cazorla16,sagotra17,sagotra18,min20}],
complex hydrides [\onlinecite{sau21}] and multiferroic perovskite oxides [\onlinecite{menendez20,machado21}]. 
This Review is addressed to both experienced computational materials scientists with an interest on 
caloric effects and solid-state cooling as well as to novice researchers wanting to learn modeling 
techniques of phase transitions and free-energy methods.

\section{Modeling of atomic interactions}
\label{sec:modelling}
The fundamentals of density functional theory (DFT), a first-principles approach widely used in condensed-matter 
and materials science, and classical interatomic potentials are briefly reviewed next. The latter interatomic 
interactions modeling approach is approximate and typically rely on the results of accurate first-principles 
methods (also called \emph{ab initio}) like DFT. On the other hand, the computational cost of DFT methods is 
several orders of magnitude higher than that of classical interatomic potentials, hence in many occasions 
first-principles methods may not be directly employed in the study of BC effects. Machine learning potentials 
are also succinctly described since they constitute a modern atomistic modeling technique in the middle way 
between DFT and classical potential approaches both in terms of accuracy and computational expense.

\subsection{$Ab$ $initio$ methods}
\label{subsec:abinitio}
In solids, the dynamics of the electrons and nuclei can be decoupled to a good approximation because 
their respective masses differ by several orders of magnitude. The wave function of the corresponding 
many-electron system, $\Psi ({\bf r}_{1}, {\bf r}_{2},...,{\bf r}_{N})$, therefore can be determined 
by solving the Schr\"{o}dinger equation corresponding to the non-relativistic Born-Oppenheimer Hamiltonian:
\begin{eqnarray}
H = -\frac{1}{2} \sum_{i} {\bf \nabla}^{2}_{i} - \sum_{I} \sum_{i} \frac{Z_{I}}{|{\bf R}_{I} - {\bf r}_{i}|} \nonumber \\
        + \frac{1}{2} \sum_{i} \sum_{j \neq i} \frac{1}{|{\bf r}_{i} - {\bf r}_{j}|}~,
\label{eq:BO-hamilton}
\end{eqnarray}
where $Z_{I}$ are the nuclear charges, ${\bf r}_{i}$ the positions of the electrons, and ${\bf R}_{I}$ 
the positions of the nuclei (considered to be fixed). In real materials, $\Psi$ is a complex 
mathematical function that in most cases is unknown. At the heart of any first-principles method is to find 
a good approximation for $\Psi$, or an equivalent quantity (e.g., the electronic density), that is manageable 
enough to perform calculations. Examples of \emph{ab initio} methods include density functional theory 
(DFT), M\o ller-Plesset perturbation theory (MP2), the coupled-cluster method with single, double and 
perturbative triple excitations [CCSD(T)], and quantum Monte Carlo (QMC), to cite just a few. Among these 
techniques, DFT methods are frequently applied to the study of materials and pressure-induced phase transitions 
hence for this reason we summarise their foundations in what follows.   

In 1965, Kohn and Sham developed a pioneering theory to effectively calculate the energy and properties 
of many-electron systems without the need of explicitly knowing $\Psi$ [\onlinecite{kohn65,sham66}]. 
The main idea underlying this theory, called density functional theory (DFT), is that the exact 
ground-state energy, $E$, and electron density, $n({\bf r})$, can be determined by solving an effective 
one-electron Schr\"odinger equation of the form:
\begin{equation}
H_{\rm KS} \Psi_{i \sigma} = \epsilon_{i \sigma} \Psi_{i \sigma}~, 
\label{eq:onelectron}
\end{equation}
where $H_{\rm KS}$ is the Kohn-Sham Hamiltonian, index $i$ labels different one-electron orbitals and 
$\sigma$ different spin states. The KS Hamiltonian can be expressed as:
\begin{equation}
H_{\rm KS} = -\frac{1}{2}\nabla^{2} + V_{ext}({\bf r}) + \int \frac{n({\bf r'})}{|{\bf r} - {\bf r'}|} d{\bf r'} + V_{xc}({\bf r})~,
\label{eq:heff}
\end{equation}
where
\begin{equation}
n({\bf r}) = \sum_{i \sigma} |\Psi_{i \sigma} ({\bf r})|^{2}~,
\label{eq:density}
\end{equation}
$V_{ext}$ represents an external field and $V_{xc} ({\bf r}) = \delta E_{xc} / \delta n ({\bf r})$
is the exchange-correlation potential.

The exchange-correlation energy has a purely quantum mechanical origin and can be defined as the
interaction energy difference between a quantum many-electron system and its classical counterpart.
Despite $E_{xc}$ represents a relatively small fraction of the total energy, this contribution is
extremely crucial for all materials and molecules because it acts directly on the bonding between
atoms. In general, $E_{xc}[n]$ is unknown and needs to be approximated. This is the only source
of fundamental and uncontrollable errors in DFT methods. In standard DFT approaches, $E_{xc}[n]$ 
typically is approximated with the expression:
\begin{equation}
E_{xc}^{\rm approx}[n] = \int \epsilon_{xc}^{\rm approx}({\bf r}) n({\bf r}) d{\bf r}~,
\label{eq:excapprox}
\end{equation}
where $\epsilon_{xc}^{\rm approx}$ is made to depend on $n({\bf r})$, $\nabla n({\bf r})$,
and/or the electronic kinetic energy $\tau ({\bf r}) = \frac{1}{2} \sum_{i \sigma} 
|\nabla \Psi_{i \sigma} ({\bf r})|^{2}$~. Next, we summarise the basic aspects of the 
most popular $E_{xc} [n]$ functionals employed in modeling of materials and phase transitions. 

In local approaches (e.g., local density approximation --LDA--), $E_{xc}^{\rm approx}$ in 
Eq.(\ref{eq:excapprox}) is calculated by considering the exchange-correlation energy of an 
uniform electron gas with density $n({\bf r})$, $\epsilon_{xc}^{unif}$, which is known exactly 
from quantum Monte Carlo calculations [\onlinecite{ceperley80,perdew81}]. In order to deal 
with the non-uniformity of real electronic systems, the space is partitioned into infinitesimal 
volume elements that are considered to be locally uniform. In semi-local approaches (e.g., 
generalized gradient approximation --GGA--), $E_{xc}$ is approximated similarly than in local 
approaches but $\epsilon_{xc}^{\rm approx}$ is made to depend also on the gradient of $n({\bf r})$  
[\onlinecite{perdew92,perdew96}]. Both local and semi-local approximations satisfy some exact 
$E_{xc}$ constraints and can work notably well for systems in which the electronic density 
varies slowly over the space (e.g., crystals). An extension of the GGA approach is provided 
by meta-GGA functionals, in which the non-interacting kinetic energy density is considered 
also as an energy functional input. An example of this latter type of functionals is the 
recently proposed meta-GGA SCAN [\onlinecite{scan,zhang18}].

Hybrid functionals comprise a combination of non-local exact Hartree-Fock and local exchange energies, 
together with semi-local correlation energies. The proportion in which both non-local and local exchange 
densities are mixed generally relies on empirical rules. The popular B3LYP approximation [\onlinecite{becke93}], 
for instance, takes a $20$\% of the exact HF exchange energy and the rest from the GGA and LDA functionals. 
Other well-known hybrid functionals are PBE0 [\onlinecite{adamo99}] and the range-separated HSE06 proposed 
by Scuseria and collaborators [\onlinecite{hse06}]. In contrast to local and semi-local functionals, 
hybrids describe to some extent the delocalisation of the exchange-correlation hole around an electron 
hence they partially correct for electronic self-interaction errors (which are ubiquitous in standard
DFT) [\onlinecite{franchini14}]. This technical feature is specially useful when dealing with strongly 
correlated systems that contain $d$ and $f$ electronic orbitals (e.g., transition-metal oxide perovskites) 
[\onlinecite{bilc08,cazorla16b,cazorla17c}].

\subsection{Classical interatomic potentials}
\label{subsec:potentials}
Using first-principles methods to describe the interactions between electrons and ions in crystals 
requires dedicated computational resources. In some cases, the interatomic interactions can be 
approximated satisfactorily by analytical functions known as classical interatomic potentials or 
force fields and consequently the simulations can be accelerated dramatically with respect to \emph{ab initio} 
calculations. Classical interaction models contain a number of parameters that are adjusted to 
reproduce experimental or \emph{ab initio} data, and their analytical expressions are constructed 
based on physical knowledge and intuition. The force matching method proposed by Ercolessi and Adams 
[\onlinecite{ercolessi94}] is an example of a classical potential fitting technique that is widely 
employed in the fields of condensed matter physics and materials science [\onlinecite{cazorla15c,mattoni15,
hata17,cazorla18b}]. Nonetheless, the ways in which classical interatomic potentials are constructed 
are neither straightforward nor uniquely defined and the thermodynamic intervals over which they 
remain reliable are limited.

A typical pairwise interaction model that has been widely employed to simulate materials at 
finite temperatures is the Coulomb--Buckingham (CB) potential, which adopts the simple form 
[\onlinecite{tinte99,jackson05,sepliarsky01,hashimoto14}]:
\begin{equation}
V_{\alpha \beta}(r_{ij}) = A_{\alpha \beta}e^{-\frac{r_{ij}}{\rho_{\alpha \beta}}} - \frac{C_{\alpha \beta}}{r_{ij}^{6}} + \frac{Z_{\alpha}Z_{\beta}}{r_{ij}}~,
\label{eq:BMH}
\end{equation}
where subscripts $\alpha$ and $\beta$ represent atomic species in the system, $r_{ij}$ the radial 
distance between a pair of $\alpha$ and $\beta$ atoms labelled $i$ and $j$ respectively, $Z$ ionic 
charges, and $A$, $\rho$ and $C$ are potential parameters. The CB potential is composed of three 
different energy contributions. The exponential term accounts for short-range repulsive forces 
resulting from the interactions between nearby electrons; the second term represents long-range 
attractive interactions arising from dispersive van der Waals forces; the third term is the usual 
Coulomb interaction between point charges. In order to describe atomic polarizability effects, 
``core-shell'' modeling can be performed on top of the CB potential. In standard core-shell approaches, 
each atom is decomposed into a charged core, which interact with others through $V_{ij}$'s analogous 
to the expression shown in Eq.(\ref{eq:BMH}), and a charged shell that is harmonically bound to the 
core [\onlinecite{jackson05,sepliarsky01,hashimoto14,gambuzzi14}].  

Some technologically relevant materials (e.g., oxide and hybrid organic-inorganic perovskites) are 
characterized by a delicate balance between short-range and long-range forces, which may be respectively 
originated by electronic orbital hybridizations and the Coulomb interactions between permanent electric 
dipoles and higher-order moments [\onlinecite{cohen92}]. The simplicity of the pairwise interaction model 
in Eq.(\ref{eq:BMH}) may not be adequate to fully grasp the complexity of such variety of interatomic 
interactions. Bond-valence (BV) potentials, for instance, represent an improvement with respect to the 
CB model because they can mimic chemical bonding in complex materials more precisely [\onlinecite{haomin17}]. 
(Many other interatomic interaction models going beyond CB exist, like the embedded atom potential 
[\onlinecite{taioli07,cazorla07}] and multibody force fields [\onlinecite{cazorla15c}], but for the sake
of focus those are not covered here.)   

A general BV potential is [\onlinecite{grinberg02,grinberg04,shin05,liu13}]:
\begin{equation}
V_{\rm BV}(r,\theta) = V_{\rm bind}(r) + V_{\rm charge}(r) + V_{\rm rep}(r) + V_{\rm nl}(\theta)~,
\label{eq:BV}
\end{equation}
where the first term in the right-hand side represents the bond-valence potential energy, the second 
the Coulomb potential energy, the third the repulsive potential energy, and the fourth an angle potential 
energy (to prevent unphysically large distortions of covalently connected structural units; for 
instance, the oxygen octahedra in oxide perovskites and molecular cations in hybrid organic-inorganic 
compounds). 

The bond-valence energy term generally is expressed as:
\begin{equation}
V_{\rm bind}(r) = \sum_{\alpha = 1}^{N_{s}} S_{\alpha} \sum_{i = 1}^{N_{\alpha}} \mid V_{i \alpha}(r_{i}) - V_{\alpha} \mid^{\gamma_{\alpha}}, 
\label{eq:BV-binding}
\end{equation}
with
\begin{equation}
V_{i \alpha}(r_{i}) = \sum_{\beta = 1}^{N_{s}} \sum_{j}^{NN} \left( \frac{r_{0}^{\alpha \beta}}{r_{ij}^{\alpha \beta}} \right)^{C_{\alpha \beta}},
\label{eq:BV-binding-II}
\end{equation}
where $N_{s}$ represents the number of atomic species in the system, $S_{\alpha}$ are fitting parameters, 
$N_{\alpha}$ the number of $\alpha$ atoms, $V_{\alpha}$ the desired atomic valence for ion $\alpha$, 
$\gamma_{\alpha}$ fitting parameters typically set to $1$, $j$ an atomic index that runs over nearest-neighbour 
(NN) ions, $r_{0}^{\alpha \beta}$ and $C_{\alpha \beta}$ parameters determined by empirical rules, and 
$r_{ij}^{\alpha \beta}$ the radial distance between ions $i$ and $j$. 

For the repulsive energy term, $V_{\rm rep}$, the following expression normally is employed:
\begin{equation}
V_{\rm rep}(r) = \epsilon \sum_{\alpha = 1}^{N_{s}} \sum_{i = 1}^{N_{\alpha}} \sum_{\beta = 1}^{N_{s}} \sum_{j = 1}^{N_{\beta}} \left( \frac{B_{\alpha \beta}}{r_{ij}^{\alpha \beta}} \right)^{12},
\label{eq:BV-rp}
\end{equation}    
where $\epsilon$ and $B_{\alpha \beta}$ are fitting parameters. Meanwhile, an harmonic function is used
for the angle potential energy that reads:
\begin{equation}
V_{\rm nl}(\theta) = k \sum_{i = 1}^{N_{\rm bun}} \left( \theta_{i,x}^{2} + \theta_{i,y}^{2} + \theta_{i,z}^{2} 
                                              \right), 
\label{eq:BV-angle}
\end{equation}
where $k$ is a fitting parameter, $N_{\rm bun}$ the number of covalently bonded units in the simulated system 
(e.g., oxygen octahedral in oxide perovskites and molecular cations in hybrid organic-inorganic compounds), 
and $\lbrace \theta_{i,\gamma} \rbrace$ the angular degrees of freedom of such entities.  

Reliable BV, or formally analogous, interatomic potentials have been developed for archetypal ferroelectric 
materials (e.g., BaTiO$_{3}$, PbTiO$_{3}$, and PbZr$_{0.2}$Ti$_{0.8}$O$_{3}$) [\onlinecite{grinberg02,grinberg04,
shin05,liu13,xu15}], hybrid organic-inorganic perovskites (e.g., MAPbI$_{3}$) [\onlinecite{mattoni15,
hata17}] and complex hydrides (e.g., Li$_{2}$B$_{12}$H$_{12}$ and LiCB$_{11}$H$_{12}$) [\onlinecite{sau21b}],
among others. Recently, the outcomes of molecular dynamics simulations performed with some of those classical 
force fields have predicted the existence of novel and giant caloric effects [\onlinecite{sau21,qi18,liu16}].

\subsection{Machine learning potentials}
\label{subsec:MLpot}
Situations in which the use of first-principles methods is prohibitive and the available classical potentials
are not versatile enough to reproduce the phase transition phenomena of interest, machine learning techniques
may be very useful. Machine learning (ML) is a subfield of artificial intelligence that exploits the systematic
identification of correlation in data sets to make predictions and analysis [\onlinecite{behler10,rupp15}].

When calculations are performed in a series of similar systems or a number of configurations involving a 
same system, the results contain redundant information. An example is to run a molecular dynamics simulation 
in which the total internal energy and atomic forces are calculated at each time step; after a sufficiently 
long time, points which are close in configurational space and have similar energies are visited during the 
sampling of the potential-energy surface. Such a redundancy can be exploited to perform computationally 
intensive calculations (i.e., of first-principles type) only in few selected configurations and then use
machine learning (ML) to interpolate between those, thus obtaining approximate solutions for most configurations. The success of this approach depends on a balance between incurred errors due to interpolation and invested 
computational effort.

ML modeling tools can provide both the energy and atomic forces directly from the atomic positions, hence 
they can be regarded as a particular class of atomistic potential. ML potentials, however, rely on very 
flexible analytic functions rather than on physically motivated functionals hence they may be highly 
versatile without requiring prior detailed knowledge of the simulated system. Promising analytic approaches 
that have been recently proposed to construct ML potentials include permutation invariant polynomials, the 
modified Shepard method using Taylor expansions, Gaussian processes and artificial neural networks. Artificial 
neural networks, in particular, have been demonstrated to be a class of ``universal approximators'' 
[\onlinecite{behler15}] since allow to approximate unknown multidimensional functions to within arbitrary 
accuracy for a given set of known function values.

ML potentials are being employed with increasing frequency to solve problems in materials science via atomistic 
simulations due to their ability to reach high accuracy levels at moderate computational expense (see work 
[\onlinecite{deringer19}] for a general review on this topic). In the context of caloric effects, only few 
works relying on ML techniques can be found in the literature (see, for instance, works [\onlinecite{zhang20,
gong22}]); however, based on the recent progression realized in other similar research fields, we foresee a 
surge in the application of ML methods to the analysis and prediction of caloric effects in the near future.

\section{Free-energy methods}
\label{sec:freener}
Barocaloric (BC) effects associated with first-order phase transitions that entail abrupt volume and structural
changes typically are large. The two main physical descriptors of BC effects are the isothermal entropy 
change, $\Delta S_{T}$, and adiabatic temperature change, $\Delta T_{S}$, associated to the phase transition. 
These two quantities are thermodynamically related, as we explain next. Consider that the entropy of a 
condensed matter system depends on temperature and pressure, $S(T,P)$; an infinitesimal entropy variation 
then can be expressed as:
\begin{equation}
dS = \left(\frac{\partial S}{\partial T}\right)_{P} dT + \left(\frac{\partial S}{\partial P}\right)_{T} dP~.
\label{eq:adiabatic1}
\end{equation}       
In an adiabatic process the entropy is conserved, $dS = 0$, thus from the equation above it follows:
\begin{equation}
\frac{C_{P}}{T}~ dT = - \left(\frac{\partial S}{\partial P}\right)_{T} dP~,
\label{eq:adiabatic2}
\end{equation}
in which the constant-pressure heat capacity $C_{P} \equiv T \left( \partial S / \partial T \right)_{P}$ has 
been introduced. 

By integrating both sides of the latter equality, one obtains:
\begin{equation}
\Delta T_{S} = - \int \frac{T}{C_{P}}~ dS_{T}~.
\label{eq:adiabatic3}
\end{equation}
which typically is approximated like [\onlinecite{cazorla19b,moya14}]:
\begin{equation}
\Delta T_{S} \approx - \frac{T_{t}}{C_{P}} \cdot \Delta S_{T}~,
\label{eq:adiabatic4}
\end{equation} 
where $T_{t}$ corresponds to the phase transition temperature and $C_{P}$ to the averaged heat capacity of the 
involved phases close to the transition point.  

Atomistic simulations generally are performed in the $(N,V,T)$ and $(N,P,T)$ ensembles hence the entropy 
and other related thermodynamic quantities can be theoretically determined as a function of the state 
variables $V$, $P$, $T$ and $\rho$ ($\equiv N/V$). Concerning BC effects, once the thermodynamic function 
$S(P,T)$ of the involved phases are known both the $\Delta S_{T}$ and $\Delta T_{S}$ descriptors can be 
straightforwardly calculated [see Eqs.(\ref{eq:adiabatic1})-(\ref{eq:adiabatic4}) above]. Meanwhile, the 
entropy of a given system is related to its Gibbs free energy, $G(P,T)$, through the expression:
\begin{equation}
S = - \left( \frac{\partial G}{\partial T} \right)_{P}~,
\label{eq:entropy}
\end{equation}        
thus it follows, without too much surprise, that the fundamental quantity allowing for the theoretical
determination of BC effects is $G(P,T)$. For this reason, in this section we explain several computational 
approaches that allow for direct estimation of free energies in crystals by using atomic interaction modeling 
techniques like those introduced in the previous section (i.e., \textit{ab initio}, classical potential and 
machine learning methods).

\subsection{The quasi-harmonic approximation}
\label{subsec:QHA}
In the quasi-harmonic approach (QHA) [\onlinecite{harmonic1,harmonic2,harmonic3}] one assumes that the 
potential energy of a crystal can be approximated with a quadratic expansion around the equilibrium atomic 
configuration of the form: 
\begin{equation} 
E_{\rm qh} = E_{\rm eq} + \frac{1}{2}
\sum_{l\kappa\alpha,l'\kappa'\alpha'}
\Phi_{l\kappa\alpha,l'\kappa'\alpha'} u_{l\kappa\alpha}
u_{l'\kappa'\alpha'}~,
\label{eq:eqh}
\end{equation}
where $E_{\rm eq}$ is the total energy of the equilibrium lattice, $\boldsymbol{\Phi}$ the corresponding 
force-constant matrix, and $u_{l\kappa\alpha}$ the displacement along Cartesian direction $\alpha$ of 
atom $\kappa$ at lattice site $l$. The atomic displacements normally are expressed as: 
\begin{equation}
u_{l\kappa\alpha}(t) = \sum_{q} u_{q\kappa\alpha} \exp{ \left[ i
    \left(\omega t - \boldsymbol{q} \cdot (\boldsymbol{l}+
    \boldsymbol{\tau}_{\kappa} \right) \right] }~,
\end{equation}
where $\boldsymbol{q}$ is a wave vector in the first Brillouin zone (BZ) that is defined by the equilibrium 
unit cell; $\boldsymbol{l}+\boldsymbol{\tau}_{\kappa}$ is the vector that locates atom $\kappa$ at cell $l$ 
in the equilibrium structure. The normal modes of the crystal then are found by diagonalizing the dynamical 
matrix:
\begin{equation}
\begin{split}
& D_{\boldsymbol{q};\kappa\alpha,\kappa'\alpha'} =\\ &
  \frac{1}{\sqrt{m_{\kappa}m_{\kappa'}}} \sum_{l'}
  \Phi_{0\kappa\alpha,l'\kappa'\alpha'} \exp{\left[
      i\boldsymbol{q}\cdot(\boldsymbol{\tau}_{\kappa}-\boldsymbol{l'}-\boldsymbol{\tau}_{\kappa'})
      \right]}~,
\end{split}
\end{equation}
and thus the solid can be treated as a collection of non-interacting harmonic oscillators with frequencies 
$\omega_{\boldsymbol{q}s}$ (positively defined and non-zero) and energy levels:
\begin{equation}
E_{\boldsymbol{q}s} = \hbar \omega_{\boldsymbol{q}s} \left( \frac{1}{2} + n \right)~,
\end{equation}
where $n$ is the Bose-Einstein occupation number. Accordingly, the potential energy of a crystal can be written 
within the QHA like: 
\begin{equation}
E_{\rm qh} = \sum_{\boldsymbol{q}s} \hbar \omega_{\boldsymbol{q}s} \left[ \frac{1}{2} + 
             \frac{1}{\exp{\left(\hbar \omega_{\boldsymbol{q}s} / k_{B}T \right)} - 1} \right] 
\label{eq:eqh2}
\end{equation}
and the constant-volume heat capacity like:
\begin{eqnarray}
& C_{V}(T) & = \left( \frac{\partial E_{\rm qh}}{\partial T} \right)_{V} \nonumber \\
&          & = \sum_{\boldsymbol{q}s} \frac{\left( \hbar \omega_{\boldsymbol{q}s}\right)^{2}}{k_{B}T^{2}}
               \frac{\exp{\left(\hbar \omega_{\boldsymbol{q}s} / k_{B}T \right)}}{\left[
               \exp{\left(\hbar \omega_{\boldsymbol{q}s} / k_{B}T \right)} - 1 \right]^{2}}~.
\label{eq:heatcap}
\end{eqnarray}
Likewise, the Helmholtz free energy, $F(V,T)$, is analytically expressed as: 
\begin{equation}
F(V,T) = E_{0}(V) + k_{B} T \sum_{\boldsymbol{q}s}\ln\left[ 2\sinh \left( 
    \frac{\hbar\omega_{\boldsymbol{q}s}(V)}{2k_{\rm B}T} \right) \right]~,
\label{eq:fharm}
\end{equation}
where $E_{0}$ corresponds to the energy of the static lattice and the dependence of the vibrational 
frequencies on volume has been explicitly noted. The Helmholtz and Gibbs free energies are related 
through the Legendre transformation $G = F + PV$, in which the pressure is equal to $P = - \left(
\partial F / \partial V \right)_{T}$. Thus, by knowing $F(V,T)$, $C_{V}(T)$ and the equation of 
state $P(V,T)$, one can deduce the value of $G$ and $C_{P}$ as a function of $P$ and $T$. 

Crystals in which anharmonic effects are important (e.g., materials in which imaginary phonon frequencies 
appear in the vibrational phonon spectra calculated at zero temperature), however, are not adequately 
described by the QHA (e.g., the Helmholtz free energy expression in Eq.(\ref{eq:fharm}) is ill-defined
and cannot be evaluated). In such a case, one should resort instead to genuine anharmonic free-energy 
approaches for the calculation of $F(V,T)$ like molecular dynamics, thermodynamic integration from 
reference models [\onlinecite{taioli07,cazorla07,cazorla12}] and self-consistent phonon approaches 
[\onlinecite{scp1,scp2,scp3,scp4}], to cite some examples. The computational load associated with 
these anharmonic free-energy methods, however, typically is orders of magnitude higher than that of 
quasi-harmonic approaches. Next, we briefly describe some of them.

\subsection{Molecular dynamics}
\label{subsec:MD}
Molecular dynamics (MD) is a computer simulation method in which the trajectories of atoms in a many-body 
interacting system are determined numerically by solving the Newton's equations of motion [\onlinecite{gulp,
lammps}]. For this purpose, one needs to know the forces acting between the particles and their potential 
energies, which can be calculated by any of the methods described in Sec.\ref{sec:modelling}. Thermostat and
barostat techniques are employed in MD simulations to render constant temperature and pressure conditions
[\onlinecite{frenkel-book}]. Temperature and anharmonic effects in materials are naturally accounted
for in MD simulations (quantum nuclear effects [\onlinecite{harmonic1}], on the other hand, are systematically 
disregarded in this approach). Estimation of quantities like $T$--renormalized phonon frequencies, $\omega 
(V,T)$, and vibrational density of states, $g(\omega)$, can be obtained from MD simulations by using 
phonon-mode decomposition [\onlinecite{dynaphopy}], force-constant optimization [\onlinecite{hiphive}] and 
effective Hamiltonian fitting [\onlinecite{hellman13}] techniques, to cite some examples.

When the interatomic interactions in MD simulations are described with first-principles methods, typically
DFT potentials, they are referred to as \emph{ab initio} molecular dynamics (AIMD). AIMD simulations certainly
are several orders of magnitude computationally more intensive than MD simulations performed with classical 
potentials. Nevertheless, thanks to the current steady growth in computational power and improved algorithms 
design (e.g., linear scaling DFT methods [\onlinecite{conquest}]), reliable AIMD simulation of complex materials 
currently is within reach (see, for instance, works [\onlinecite{sagotra19,miyazaki21,gebbia21,polek20,luo20}]). 
Machine learning based molecular dynamics, on the other hand, represents an excellent alternative to AIMD 
simulations (provided that a reliable interaction model can be generated) due to its great balance between 
numerical accuracy and computational expense [\onlinecite{jinnouchi19}]. 

A standard approach to estimate $T$--renormalized $g(\omega)$'s from MD simulations consists in calculating 
the Fourier transform of the velocity-velocity autocorrelation function as follows 
[\onlinecite{sagotra19,miyazaki21,gebbia21}]:
\begin{equation}
        g(\omega) = \frac{1}{N_{ion}} \sum_{i}^{N_{ion}} \int_{0}^{\infty} 
        \langle {\bf v}_{i}(t)\cdot{\bf v}_{i}(0)\rangle e^{i\omega t} dt~,
\label{eq5}     
\end{equation}
where ${\bf v}_{i}(t)$ represents the velocity of the atom labelled as $i$ at time $t$, and $\langle \cdots 
\rangle$ denotes statistical average in the $(N,V,T)$ ensemble.
By knowing g($\omega$), one can easily estimate $F(V,T)$ from the integral expression that is equivalent 
to Eq.(\ref{eq:fharm}), namely:
\begin{equation}
F(V,T) = E_{0}(V) + \int_{0}^{\infty} k_{B} T \ln\left[ 2\sinh \left(\frac{\hbar\omega}{2k_{\rm B}T} \right) \right] g(\omega) d\omega~.
\label{eq:fharm2}
\end{equation} 
Likewise, the constant-volume heat capacity can be computed by replacing the summation in Eq.(\ref{eq:heatcap})
with the corresponding integral expression, and the entropy with the formula [\onlinecite{phonopy}]:
\begin{eqnarray}
& S(V,T) & = - \left( \frac{\partial F}{\partial T} \right)_{V} = \int_{0}^{\infty} \frac{\hbar\omega}{2T} \coth \left(\frac{\hbar\omega}{2k_{\rm B}T} \right) - \nonumber \\
&       &  k_{B}  \ln\left[ 2\sinh \left(\frac{\hbar\omega}{2k_{\rm B}T} \right) \right] g(\omega) d\omega~. 
\label{eq:entropy2}
\end{eqnarray}  
Thus, just like in the QHA method, one can deduce the value of the thermodynamic functions $G(P,T)$ and 
$C_{P}(T)$, and in turn of the barocaloric descriptors $\Delta S_{T}$ and $\Delta T_{S}$, from the knowledge 
of $F(V,T)$, $C_{V}(T)$ and $P(V,T)$.

\subsection{Thermodynamic integration}
\label{subsec:thermoint}
More often than not, accurate estimation of the Helmholtz free energy within an ample thermodynamic 
interval based on Eqs.(\ref{eq5})--(\ref{eq:fharm2}) may turn out to be computationally too intensive. In such 
a situation, thermodynamic integration techniques, in their different variants, may result very useful since 
they rely on the fact that thermodynamic quantities other than $F(V,T)$ (e.g., the internal energy and pressure) 
can be numerically converged more efficiently in practice. Two popular modalities of thermodynamic integration 
(TI) are (1)~standard TI and (2)~TI from reference models [\onlinecite{taioli07,cazorla07,cazorla12,cazorla09}]. 

Standard TI exploits two well-known thermodynamic relations. First, by considering the hydrostatic 
pressure expression:
\begin{equation}
\left(\frac{\partial F}{\partial V} \right)_{V} = -P~,
\label{eq:pressure}
\end{equation}  
it is possible to estimate Helmholtz free energy shifts as a function of volume at a fixed temperature like:
\begin{equation}
F(V_{1},T) - F(V_{0}, T) = \int_{V_{1}}^{V_{0}} P ~dV~.
\label{eq:sti1}
\end{equation}
Likewise, based on the following expression deduced from statistical mechanics:
\begin{equation}
\left( \frac{\partial \left[ \beta F \right]}{\partial \beta} \right)_{V} = E~,
\label{eq:statistical}
\end{equation}
where $\beta \equiv 1/k_{B}T$, it is possible to estimate Helmholtz free energy shifts as a function of 
temperature at a fixed volume like:
\begin{equation}
\beta_{1}F(V,T_{1}) - \beta_{0}F(V, T_{0}) = \int_{\beta_{0}}^{\beta_{1}} E ~d\beta~.
\label{eq:sti2}
\end{equation}
In view of Eqs.(\ref{eq:sti1})--(\ref{eq:sti2}), it is possible then to extend the value of the Helmholtz free 
energy by performing MD simulations over a dense grid of $(V,T)$ points, from which $P$ and $E$ can be 
determined, and subsequently compute the value of the involved integrals numerically.   

In the standard TI approach, however, only Helmholtz free energy differences can be estimated hence for
the computation of total free energies of different phases, and therefore of barocaloric effects, the 
value of $F(V,T)$ should be known at least at one thermodynamic state. TI from a reference model can be 
very useful for this end. The general principle in this approach is that the change of Helmholtz free 
energy is computed as the total energy function $E_\lambda ( {\bf r}_1 , \ldots {\bf r}_N )$ changes 
adiabatically from $E_0$ to $E_1$, the free energies associated with these energy functions being $F_0$ 
and $F_1$. Specifically, the formula for TI from a reference model is [\onlinecite{cazorla12,cazorla09}]:
\begin{equation}
F_1 - F_0 = \int_0^1 \langle E_1 - E_0 \rangle_\lambda~,
\label{eq:tifrm}
\end{equation}
where $\langle \cdots \rangle_\lambda$ represents the thermal average evaluated for the system governed by 
the energy function $E_\lambda = ( 1 - \lambda ) E_0 + \lambda E_1$. In practice, one takes $E_1$ to be 
the total energy function whose free energy wish to calculate (e.g., a specific DFT functional), and $E_0$  
the total energy function of a ``reference'' system whose free energy can be evaluated exactly. For solids, 
one typically chooses the reference system to be a perfectly harmonic system, the corresponding Helmholtz 
free energy expression being that in Eqs.(\ref{eq:fharm})--(\ref{eq:fharm2}). Meanwhile, the integral in 
Eq.~(\ref{eq:tifrm}) should be computed numerically and for this end one can perform a series of MD 
simulations in the $(N,V,T)$ ensemble governed by the energy function $E_{\lambda}$ at different $\lambda$ 
values.

\section{Estimation of BC effects}
\label{sec:estimation}
In analogy to the experiments, BC effects can be estimated with atomistic simulations in an indirect, 
quasi-direct and direct fashion. In the indirect approach, one exploits the Maxwell relation:
\begin{equation}
-\left(\frac{\partial S}{\partial P}\right)_{T} = \frac{\partial^{2} G}{\partial T \partial P} = 
 \left(\frac{\partial V}{\partial T}\right)_{P}~,
\label{eq:Maxwell}
\end{equation}
to estimate the adiabatic temperature and isothermal entropy shifts like [see Eq.(\ref{eq:adiabatic2})]:
\begin{eqnarray}
\Delta T_{S} = \int \frac{T}{C_{P}}~ \left(\frac{\partial V}{\partial T}\right)_{P}~ dP \nonumber \\
\Delta S_{T} = - \int \left(\frac{\partial V}{\partial T}\right)_{P}~ dP~.
\label{eq:indirect}
\end{eqnarray}
The temperature derivatives in Eq.(\ref{eq:indirect}), however, are ill-defined for the case of first-order 
phase transitions since the volume of the system changes discontinously with temperature during this type  
of transformation. Accordingly, indirect approaches based on the Maxwell relation above are only valid for 
estimating BC effects associated with second-order like phase transitions in which the volume of the system 
changes continuously rather than abruptly [\onlinecite{cazorla16,sagotra17,sagotra18,min20}].   

As it was mentioned previously, large BC effects typically are associated with first-order phase transitions. 
In such a case, one can approximately estimate the size of $\Delta S_{T}$ and $\Delta T_{S}$ with the 
Clausius-Clapeyron method or exactly with quasi-direct and direct techniques. We briefly describe each of 
these approaches in the following sections.

\subsection{The Clausius-Clapeyron method}
\label{sec:clausius}
To a first approximation, the barocaloric isothermal entropy change associated with a first-order 
$P$-induced phase transition can be estimated with the Clausius-Clapeyron relation like [\onlinecite{moya14}]:
\begin{equation}
\Delta S_{T} \approx \Delta S = \Delta V \cdot \frac{d P_{t}}{d T_{t}}~, 
\label{eq:clausius}
\end{equation}
where $\Delta V$ is the volume change that the system experiences during the phase transformation occurring
at conditions $P_{t}(T_{t})$ and the derivative is estimated on the corresponding phase boundary line. 
Meanwhile, the corresponding adiabatic temperature change can be roughly determined with 
Eq.(\ref{eq:adiabatic4}).  

The Clausius-Clapeyron (CC) method is not exact and entails few technical problems in practice. First, in 
experiments the entropy change of the phase transformation, $\Delta S$, is not necessarily equal to the BC 
descriptor $\Delta S_{T}$ due to the presence of pervasive hysteresis and phase coexistence effects 
[\onlinecite{cazorla19b}], which in general are not reproducible with atomistic simulation techniques (see
next Sec.\ref{subsec:quasi-direct}). And second, the value of the derivative and volume change entering 
Eq.(\ref{eq:clausius}) typically are accompanied by large numerical uncertainties that are propagated into 
the evaluation of $\Delta S$. Having said this, the CC method represents a fair approach to estimate BC 
effects and it has been already employed in numerous experimental and theoretical works 
[\onlinecite{sau21,li19,bermudez17}]. In order to estimate BC effects with the highest accuracy, however, 
it is recommended to employ quasi-direct and direct methods.

\begin{figure}[t]
\centerline
        {\includegraphics[width=1.00\linewidth]{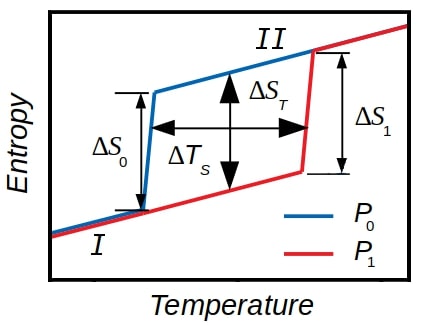}}
\caption{Sketch of the entropy curves $S(P,T)$ that can be theoretically estimated with computational
         free-energy methods for a system undergoing a pressure-induced first-order phase transition,
         $I \to II$. Hysteresis effects have been disregarded. The BC descriptors $\Delta S_{T}$ and 
         $\Delta T_{S}$ associated with the phase transition can be straightforwardly deduced from these 
         curves. The entropy changes associated with the phase transition, $\Delta S$'s, are not necessarily 
         equal to the BC isothermal entropy change $\Delta S_{T}$.}
\label{fig1}
\end{figure}

\subsection{Quasi-direct estimation of BC effects}
\label{subsec:quasi-direct}
The theoretical quasi-direct method emulates the analogous experimental quasi-direct approach (see, for 
instance, works [\onlinecite{cazorla17b,lloveras19,lloveras20}]) in which essentially the pressure 
and temperature dependences of the system entropy, $S(P,T)$, are determined (Fig.\ref{fig1}). Upon 
derivation of the full entropy curves (i.e., exhibiting a discontinuity, $\Delta S$, at the first-order 
phase transition points) obtained at constant pressures $P_{0}$ and $P_{1}$ and expressed as a function of 
temperature, it is straightforward to graphically deduce the size of the BC descriptors $\Delta S_{T}$ 
and $\Delta T_{S}$ on the corresponding $S$--$T$ diagram, as it is illustrated in Fig.\ref{fig1}. From 
a practical point of view, the $S(P,T)$ curves can be calculated numerically with any of the free-energy 
methods described in Sec.\ref{sec:freener}, and the phase transition points be determined under the condition
$G_{I}(P,T_{t}) = G_{II} (P,T_{t})$ ($I$ and $II$ being the two phases involved in the first-order 
transformation). 

It is worth mentioning, however, that there are important differences between the theoretical and
experimental quasi-direct BC approaches. In experiments, the full entropy curves, typically determined
with calorimetric techniques, inevitably include hysteresis and phase coexistence effects. Hysteresis,
for instance, originates from the fact that in practice transition paths occur under nonequilibrium 
conditions; consequently, irreversible processes take place within the material. Irreversibility may pose 
severe limitations to the overall cooling performance attained during successive field-induced cycles, 
hence it is an important aspect to take into consideration when envisaging possible solid-state cooling 
applications [\onlinecite{manosa17}]. Unfortunately, hysteresis, and in general any nonequilibrium process, 
cannot be reproduced with the simulation techniques reviewed in Sec.\ref{sec:freener} since perfect 
equilibrium and reversibility conditions are always assumed in them. Thus, in most cases the value of the
theoretically estimated BC descriptors should be regarded as upper bounds of the real $\Delta S_{T}$ and 
$\Delta T_{S}$ shifts (i.e., those measured in the experiments).

\subsection{Direct estimation of BC effects}
\label{subsec:direct}
In this simulation approach, the system is first thermalized at the desired initial temperature and 
pressure in the $(N,P,T)$, ensemble. After thermalization, the simulation is switched to the 
isobaric-isoenthalpic ensemble, $(N,P,H)$ [where $H$ represents the enthalpy of the system]. At this stage 
the hydrostatic pressure is ramped up to the desired final value slowly enough to guarantee adiabaticity, 
and the accompanying temperature change, $\Delta T_{\rm on}$, is monitored. The system is simulated under 
these conditions for some time. Subsequently, $P$ is ramped down to its initial value slowly enough again 
to guarantee adiabaticity and the corresponding temperature change, $\Delta T_{\rm off}$, and final 
temperature, $T_{f}$, are monitored. Under the condition that the system remains in thermal equilibrium 
during the entire described simulation cycle, it will follow that $T_{i} = T_{f}$ and $\Delta T_{\rm on} 
= \Delta T_{\rm off}$ within the corresponding statistical uncertainties. In such a case, either $\Delta 
T_{\rm on}$ or $\Delta T_{\rm off}$ can be identified with the adiabatic temperature change associated 
with the simulated BC process. 

The main advantages of using the direct BC simulation method are the little supervision required and 
possibility to combine pressure with additional external fields (e.g., electric bias) in order to explore 
original multicaloric effects [\onlinecite{cazorla19b}]. The disadvantages, on the other hand, are large 
computational expense, no direct access to the isothermal entropy change, and the fact that implementation 
of the $(N,P,H)$ ensemble is rare in most molecular dynamics software packages. To the best of our knowledge, 
direct estimation of BC effects has been only performed for the hybrid organic-inorganic perovskite 
CH$_{3}$NH$_{3}$PbI$_{3}$ [\onlinecite{liu16}], also known as MAPI.

\section{Representative examples}
\label{sec:examples}
In this section, we analyze several illustrative cases in which original BC effects have been simulated
and quantified by using atomistic simulation methods. The systems for which those BC effects have been
predicted are technologically relevant, namely, fast-ion conductors (LiN$_{3}$), orientationally 
disordered complex hydrides (Li$_{2}$B$_{12}$H$_{12}$) and multiferroic perovskite oxides 
(BiFe$_{1-x}$Co$_{x}$O$_{3}$). In these cases, the involved materials modeling techniques comprised molecular 
dynamics relying on classical force fields (LiN$_{3}$ and Li$_{2}$B$_{12}$H$_{12}$) and the quasi-harmonic 
approximation performed with DFT methods (BiFe$_{1-x}$Co$_{x}$O$_{3}$). Analogous BC simulation success can 
be achieved also for other families of pressure responsive materials. 

\begin{figure*}[t]
\centerline
        {\includegraphics[width=1.00\linewidth]{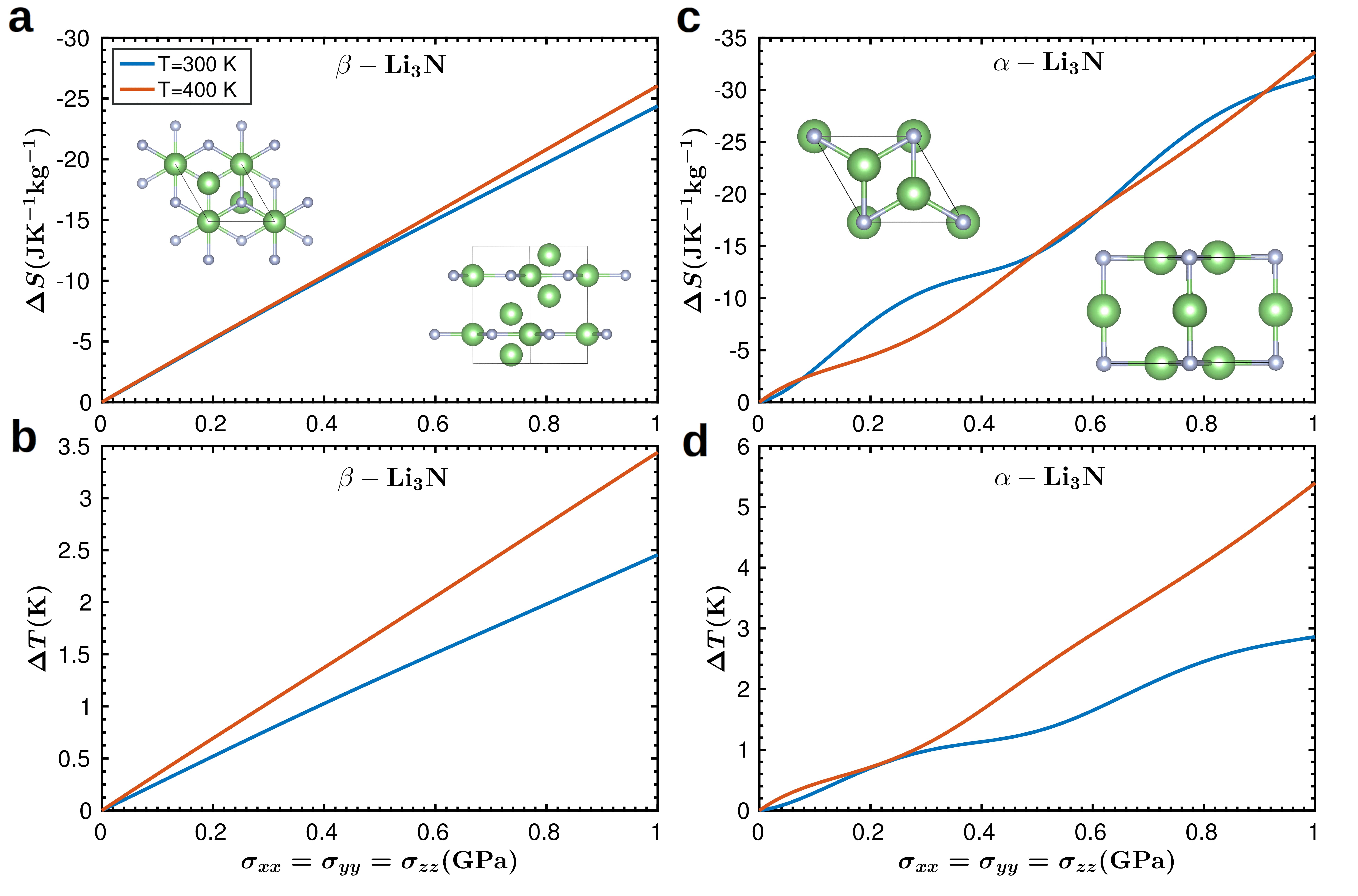}}
\caption{Predicted barocaloric effects in bulk Li$_{3}$N.
        (a)~Isothermal entropy and (b)~adiabatic temperature shifts estimated for $\beta$-Li$_{3}$N.
        (c)~Isothermal entropy and (d)~adiabatic temperature shifts estimated for $\alpha$-Li$_{3}$N.
        Ball-stick representation of the $\beta$ and $\alpha$ polymorphs of Li$_{3}$N; Li and N ions 
        are represented with green and blue spheres, respectively. Adapted from work [\onlinecite{sagotra18}].}
\label{fig2}
\end{figure*}

\subsection{Fast-ion conductor LiN$_{3}$}
\label{subsec:fic}
Fast-ion conductors (FIC) are materials that exhibit high ionic conductivity in the solid phase 
[\onlinecite{hull04}]. Examples of archetypal FIC, also known as \emph{superionic} conductors, include 
alkali-earth metal fluorides (CaF$_{2}$), oxides (doped CeO$_{2}$), and lithium-rich compounds 
(Li$_{10}$GeP$_{2}$S$_{12}$). Inherent to superionicity is a significant increase in the concentration of 
point defects (e.g., Frenkel pair defects) associated to a particular sublattice of atoms in the crystal. 
Due to their unique ion-transport properties, FIC are promising materials for the realization of all-solid-state 
electrochemical batteries via replacement of customary liquid electrolytes, which involves substantial 
improvements with respect to standard batteries in terms of safety, cyclability, and electrochemical 
performance [\onlinecite{wang15}]. Owing to the large entropy change (typically of the order of $100$~J~K$^{-1}$ 
[\onlinecite{hull04}]) and external tunability associated with the superionic phase transition 
[\onlinecite{cazorla18b,cazorla14,cazorla17d}], FIC also are being established as auspicious caloric materials 
for exploitation in solid-state cooling applications 
[\onlinecite{cazorla17b,cazorla16,sagotra18,min20,cazorla15}].

Figure~\ref{fig2} shows the two common polymorphs of bulk Li$_{3}$N, a structurally simple and well-known 
lithium-based FIC. The $\alpha$ phase (hexagonal, space group $P6/mmm$) has a layered structure composed of 
alternating planes of hexagonal Li$_{2}$N and pure Li$^{+}$ ions. The $\beta$ phase (hexagonal, space group 
$P6_{3}/mmmc$) presents an additional layer of lithium ions intercalated between the Li$_{2}$N planes that is 
accompanied by a doubling of the unit cell. Exceptionally high ionic conductivities of the order of 
$10^{-4}$--$10^{-3}$~S~cm$^{-1}$ have been experimentally observed in Li$_{3}$N at room temperature 
\cite{li10,alpen77,nazri94}. In molecular dynamics simulation studies based on classical interaction potentials 
and first-principles methods [\onlinecite{sagotra18}], it has been found that in order to reproduce 
room-temperature superionicity in $\beta$--Li$_{3}$N it is necessary to introduce a small concentration of 
extrinsic Li$^{+}$ vacancies in the system ($\sim 1$\%); by contrast, stoichiometric $\alpha$--Li$_{3}$N 
already displays fast-ionic conductivity at $T = 300$~K.

Figure~\ref{fig2} also encloses the theoretical estimation of BC effects in Li$_{3}$N up to $1$~GPa at room 
temperature and $T = 400$~K [\onlinecite{sagotra18}]. Those calculations relied on molecular dynamics simulations
performed with classical interatomic potentials (Sec.\ref{subsec:potentials}) and the set of equations 
Eq.(\ref{eq:indirect}). In the $\beta$ phase, a maximum room-temperature entropy shift of 
$-24$~JK$^{-1}$kg$^{-1}$ was obtained, which slightly increases in absolute value at higher temperatures 
(Fig.\ref{fig2}a). The magnitude of such $\Delta S$ can be regarded as ``giant'' (i.e., $|\Delta S| > 
10$~JK$^{-1}$kg$^{-1}$). The accompanying adiabatic temperature shifts, however, are large but not giant 
(that is, $|\Delta T| < 10$~K, Fig.\ref{fig2}b) owing to the huge heat capacity of Li$_{3}$N 
(e.g., $\sim 4 \cdot 10^{3}$~JK$^{-1}$kg$^{-1}$ at $T = 300$~K). Analogous BC results were found in 
$\alpha$--Li$_{3}$N (Figs.\ref{fig2}c--d) although in this latter case the estimated isothermal entropy and 
adabatic temperature changes were slightly larger (for instance, $\Delta S = -32$~JK$^{-1}$kg$^{-1}$ and 
$\Delta T = +2.8$~K at $T = 300$~K and $\sigma = 1$~GPa).

The large BC effects found in superionic Li$_{3}$N are not driven by any structural phase transformation. 
This assertion is confirmed, for instance, by the radial distribution functions calculated for all pairs of 
ionic species in Li$_{3}$N at $T = 300$~K since they present almost identical structural traits independently 
of the applied pressure [\onlinecite{sagotra18}]. Such an absence of phase transformation is in stark 
contrast to what is observed in other BC materials in which most of the caloric response is concentrated 
near phase transition points [\onlinecite{moya14}]. Then, which is the principal mechanism behind the 
disclosed giant $\Delta S$ in Li$_{3}$N? It is well-known that hydrostatic pressure depletes significantly 
ionic diffusivity, and therefore the entropy, in most fast-ion conductors [\onlinecite{cazorla18b,hull04,
cazorla14,cazorla17d}]. Essentially, the available volume to interstitial ions is effectively reduced under 
compression and as a consequence the kinetic barriers and formation energy of defects governing ion migration 
drastically increase. Thus, increasing compression steadily depletes ionic diffusion and in turn the entropy
of the FIC crystal.   

Therefore, in work [\onlinecite{sagotra18}] and original and rational strategy for achieving large (and possibly 
also reversible) BC effects at ambient temperature based on fast-ion conductors was introduced and illustrated 
for bulk Li$_{3}$N. Instead of focusing on the triggering of a structural phase transition, which only 
serendipitously will occur at room temperature, the starting point in this new scenario is a material that is 
already superionic at ambient conditions. Large BC effects then can be obtained trough the application of 
hydrostatic pressure since this affects significantly and sustainedly the ionic conductivity, and in turn the 
entropy and volume, of fast-ion conductors.

\begin{figure*}[t]
\centerline
        {\includegraphics[width=1.00\linewidth]{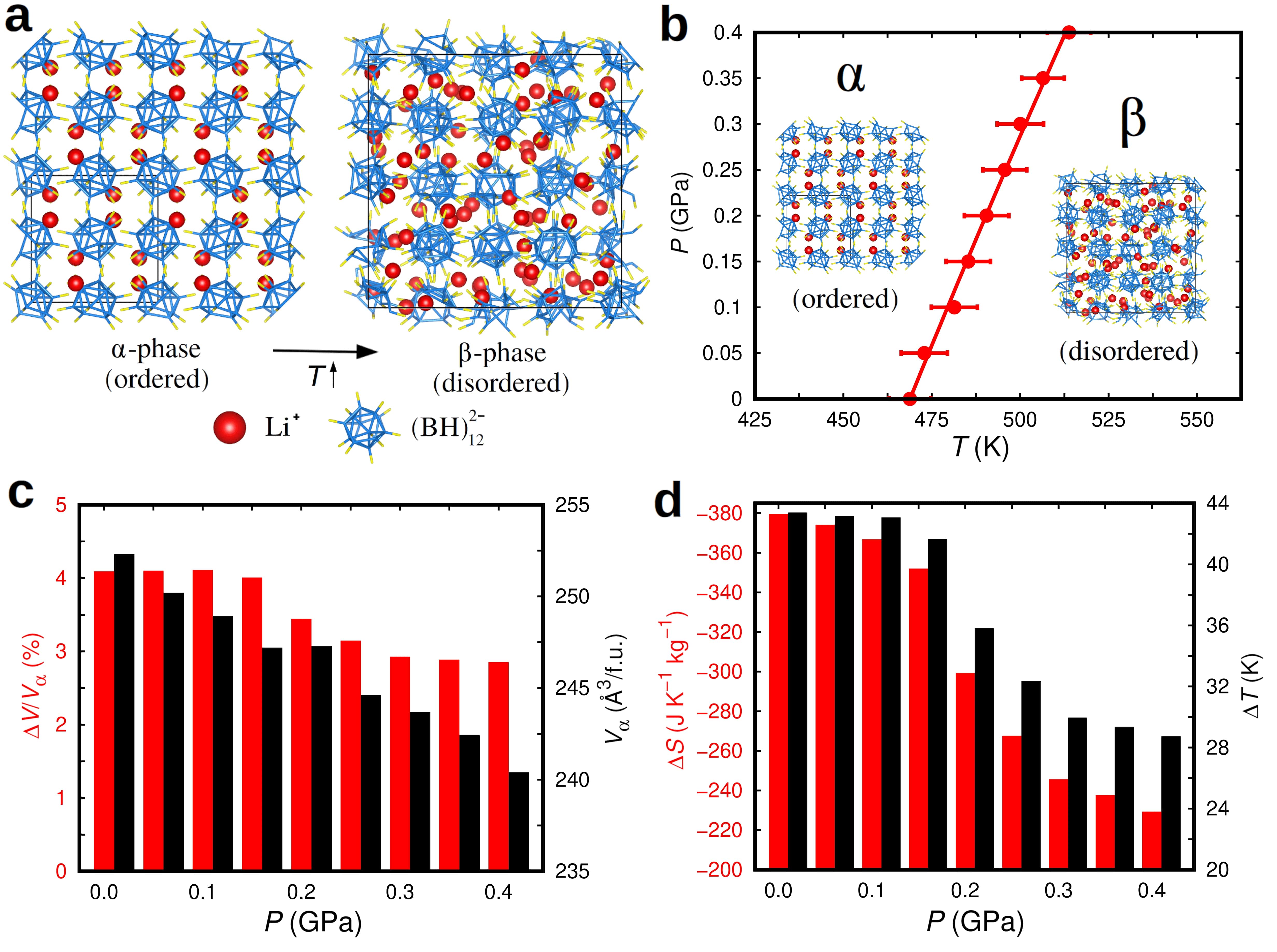}}
\caption{(a)~Low-$T$ (ordered) and high-$T$ (disordered) phases of bulk Li$_{2}$B$_{12}$H$_{12}$.
         Li, B, and H ions are represented with red, blue, and yellow colours, respectively.
         (b)~Estimated Li$_{2}$B$_{12}$H$_{12}$ phase diagram expressed as a function of pressure
         and temperature.
         (c)~Relative volume change associated with the $T$-induced $\alpha \to \beta$ phase transition 
         expressed as a function of pressure (red); volume of the $\alpha$ phase per formula unit (f.u.), 
         $V_{\alpha}$, at the phase-transition conditions (black).
         (d)~Isothermal entropy (red) and adiabatic temperature (black) changes associated with the 
         barocaloric response of bulk Li$_{2}$B$_{12}$H$_{12}$ expressed as a function of pressure.
         Adapted from work [\onlinecite{sau21}].
         }
\label{fig3}
\end{figure*}

\subsection{Complex hydride Li$_{2}$B$_{12}$H$_{12}$}
\label{subsec:LBH}
Recently, colossal barocaloric effects (defined as $|\Delta S| \sim 100$~JK$^{-1}$kg$^{-1}$) have been 
measured in two different families of materials that display intriguing order-disorder phase transitions
[\onlinecite{cazorla17b,li19,lloveras19}]. First, giant barocaloric effects have been theoretically predicted
[\onlinecite{sagotra17}] and experimentally observed in the archetypal superionic compound AgI 
[\onlinecite{cazorla17b}]. AgI exhibits a first-order normal (low-entropy) to superionic (high-entropy) phase 
transition that responds to both temperature and pressure and which involves the presence of highly mobile 
silver ions in the high--$T$ superionic state [\onlinecite{hull04}]. Likewise, the entropy changes estimated 
for other similar superionic phase transitions in general are also large 
[\onlinecite{cazorla16,sagotra18,min20}]. And second, colossal barocaloric effects have been reported for the 
molecular solid neopentylglycol [\onlinecite{li19,lloveras19}], (CH$_{3}$)$_{2}$C(CH$_{2}$OH)$_{2}$, and other 
plastic crystals [\onlinecite{lloveras20}]. In these solids, molecules reorient almost freely around their 
centers of mass, which remain localized at well-defined lattice positions. Molecular rotations lead to 
orientational disorder, which renders high entropy. By using hydrostatic pressure, it is possible to block such 
molecular reorientational motion and thus induce a fully ordered state characterized by low entropy 
[\onlinecite{cazorla19a}]. The barocaloric effects resulting from this class of first-order order-disorder 
phase transition are huge and comparable in magnitude to those achieved in conventional refrigerators with 
environmentally harmful gases [\onlinecite{li19,lloveras19,lloveras20}].

In work [\onlinecite{sau21}], it has been predicted the occurrence of colossal barocaloric effects ($|\Delta S| 
\sim 100$~JK$^{-1}$kg$^{-1}$) in the energy material Li$_{2}$B$_{12}$H$_{12}$ (LBH), a complex hydride that is 
already known from the fields of hydrogen storage [\onlinecite{her08,lai19,shevlin12}] and solid-state batteries 
[\onlinecite{paskevicius13,mohtadi16}]. By using molecular dynamics simulations [\onlinecite{sau19}], a 
pressure-induced isothermal entropy change of $|\Delta S| = 367$~JK$^{-1}$kg$^{-1}$ and an adiabatic temperature 
change of $|\Delta T| = 43$~K at $T = 480$~K have been identified. These colossal entropy and temperature 
changes were driven by small hydrostatic pressure shifts of $\sim 0.1$~GPa (Fig.\ref{fig3}), thus yielding 
gigantic barocaloric strengths of $|\Delta S| / P \sim 10^{3}$~JK$^{-1}$kg$^{-1}$GPa$^{-1}$ and $|\Delta T| / 
P \sim 10^{2}$~K~GPa$^{-1}$. The colossal barocaloric effects disclosed in bulk LBH are originated by 
simultaneous frustration of Li$^{+}$ diffusion and (BH)$_{12}^{-2}$ icosahedra reorientational motion driven 
by hydrostatic pressure. 

Specifically, at ambient conditions lithium dodecahydrododecaborate (Li$_{2}$B$_{12}$H$_{12}$), LBH, presents 
an ordered cubic $Pa\overline{3}$ phase ($Z = 4$), referred to as $\alpha$, which is characterized by Li$^{+}$
cations residing on near-trigonal-planar sites surrounded by three (BH)$_{12}^{-2}$ icosahedron anions. In
turn, each (BH)$_{12}^{-2}$ anion resides in an octahedral cage surrounded by six Li$^{+}$ cations 
(Fig.\ref{fig3}a). A symmetry preserving order-disorder phase transition occurs at high temperatures
($\sim 600$~K) that stabilises a disordered state, referred to as $\beta$, in which the Li$^{+}$
cations are mobile and the (BH)$_{12}^{-2}$ anions present reorientational motion (Fig.\ref{fig3}a). The 
relative volume expansion that has been experimentally measured for such a first-order order-disorder phase 
transition is $\Delta V^{\rm expt} / V_{\alpha}^{\rm expt} \approx 8$\% [\onlinecite{paskevicius13}] and the 
corresponding phase transition enthalpy $\Delta H^{\rm expt} \approx 130$~kJ~kg$^{-1}$ [\onlinecite{verdal14}]. 
The huge volume variation and enthalpy associated with the $\alpha \leftrightarrow \beta$ transformation could 
be propitious for barocaloric purposes if the involved phase transition was responsive to small hydrostatic 
pressure shifts of $\sim 0.1$~GPa. 

Figure~\ref{fig3}b shows the $P$--$T$ phase diagram of bulk LBH theoretically estimated with MD techniques
[\onlinecite{sau19}]. The coexistence line of the $\alpha$ and $\beta$ phases was determined by conducting 
numerous MD simulations at small $P$--$T$ shifts of $0.05$~GPa and $12.5$~K, and by monitoring the structural, 
Li$^{+}$ diffusion, and (BH)$_{12}^{-2}$ reorientational properties of the system. In consistency with the 
experiments, a point in the $\alpha$--$\beta$ coexistence line of Fig.\ref{fig3}b was identified with sharp 
and simultaneous changes in the volume (Fig.\ref{fig3}c), Li$^{+}$ diffusion coefficient ($D_{\rm Li}$), and 
(BH)$_{12}^{2-}$ reorientational frequency ($\lambda_{\rm B_{12}H_{12}}$) of bulk LBH. It was found that the 
critical temperature of the $\alpha \leftrightarrow \beta$ transformation can be certainly displaced by 
hydrostatic pressure (i.e., roughly by $13$~K per $0.1$~GPa).

Consequently, the BC isothermal entropy and adiabatic temperature shifts induced by pressures of $0 \le P \le 
0.4$~GPa in bulk LBH were theoretically estimated in work [\onlinecite{sau21}]. For this end, and in view of 
the first-order nature of the $\alpha \leftrightarrow \beta$ phase transformation \cite{paskevicius13}, the 
Clausius-Clapeyron method was employed (Sec.\ref{sec:clausius}). The resulting $\Delta S$ and $\Delta T$ values 
enclosed in Fig.~\ref{fig3}d in fact render colossal barocaloric effects. For example, at $T = 480$~K and 
$P = 0.1$~GPa an isothermal entropy change of $-367$~JK$^{-1}$kg$^{-1}$ and an adiabatic temperature change of 
$+43$~K were estimated. The size of $\Delta S$ and $\Delta T$ were found to gradually decrease under
pressure (e.g., at $T = 515$~K and $P = 0.4$~GPa we estimated $-229$~JK$^{-1}$kg$^{-1}$ and $+28$~K, 
respectively). Meanwhile, the predicted LBH barocaloric effects were always direct, that is, $\Delta T > 0$, 
since the low-entropy ordered phase ($\alpha$) was stabilized over the high-entropy disordered phase ($\beta$) 
under pressure ($\Delta S < 0$).

The phase transition underlying the colossal barocaloric effects enclosed in Fig.\ref{fig3} is quite remarkable 
since it combines key ingredients of fast-ion conductors (i.e., ionic diffusion) and molecular crystals (i.e., 
reorientational motion), materials that individually have been proven to be excellent BC materials. Thus, 
alkali-metal complex borohydrides ($A_{2}$B$_{12}$H$_{12}$, $A =$ Li, Na, K, Cs [\onlinecite{udovic14,udovic20}])
emerge as a promising new family of caloric materials in which the salient phase-transition features of fast-ion 
conductors and plastic crystals coexist and cooperate to render colossal barocaloric effects, broadening so
the range of caloric materials with commercial potential for solid-state cooling applications.

\begin{figure*}[t]
\centerline
        {\includegraphics[width=1.00\linewidth]{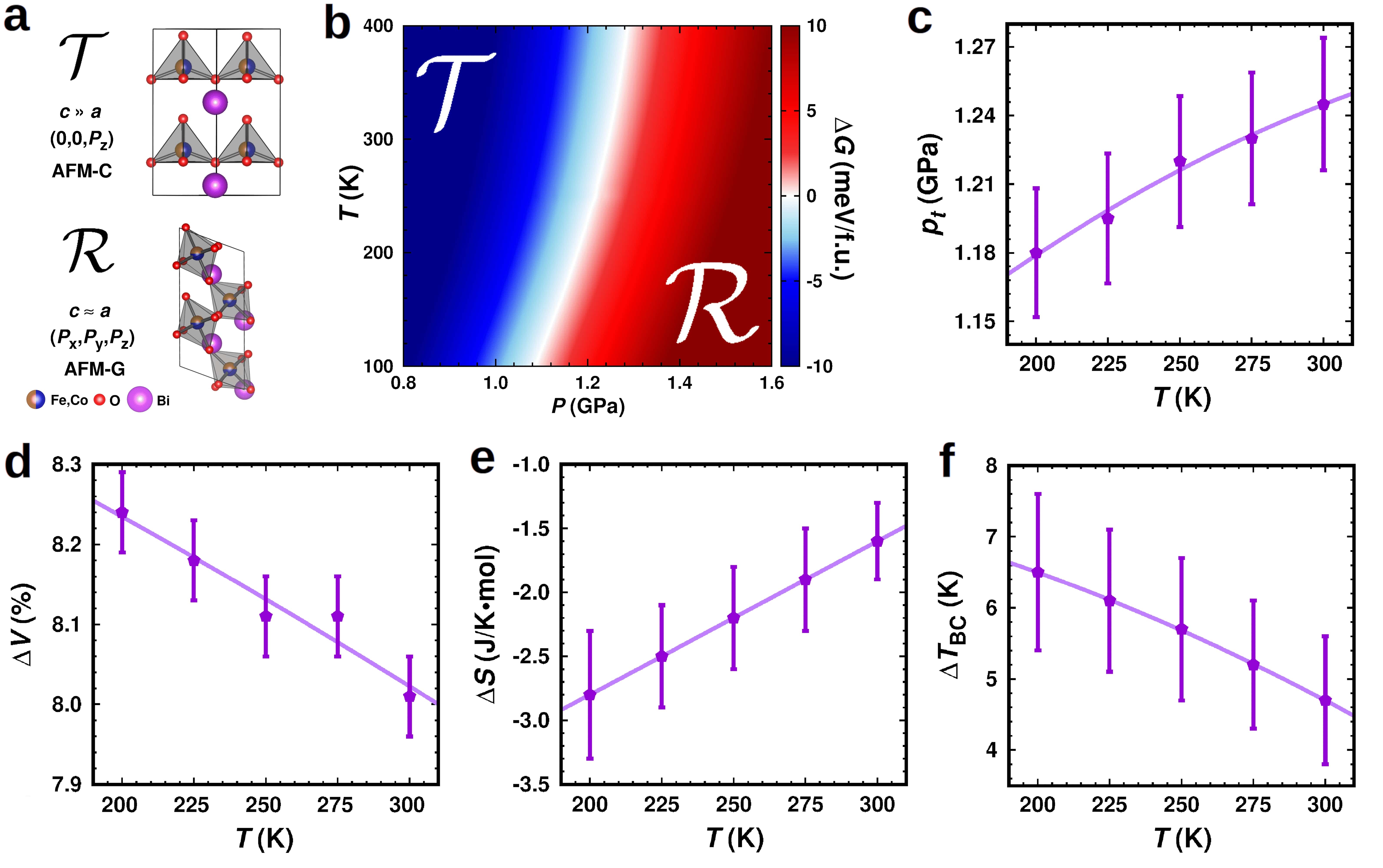}}
\caption{(a)~Description of the two energetically competing structures in bulk BFCO under pressure. 
         (b)~Gibbs free energy difference between the ${\cal T}$ and ${\cal R}$ phases of bulk 
             BFCO expressed as a function of pressure and temperature.
         (c)~Phase boundary delimiting the stability region of the ${\cal T}$ and ${\cal R}$ phases of bulk 
             BFCO.
         (d)~Relative volume change occurring during the pressure-induced ${\cal T} \to {\cal R}$ phase 
             transition and expressed as a function of temperature.
         (e)~Isothermal entropy and (f)~adiabatic temperature changes estimated for the pressure-induced 
             ${\cal T} \to {\cal R}$ phase transition expressed as a function of temperature. 
        }
\label{fig4}
\end{figure*}

\subsection{Supertetragonal oxide perovskite BiFe$_{1-x}$Co$_{x}$O$_{3}$}
\label{subsec:superT}
Super-tetragonal (${\cal T}$) oxide perovskites comprise a family of materials that are fundamentally
intriguing and have great potential for ferroelectric, piezoelectric, sensor, and energy conversion 
applications [\onlinecite{zhang18b,yamada13,infante11}]. Super-tetragonal phases exhibit giant electric 
polarizations of the order of $100$~$\mu$C/cm$^{2}$ and may be accompanied by magnetism. The coexistence 
of ferroelectricity and magnetism in crystals, known as multiferroics, offers the possibility of controlling 
the magnetization with electric fields via their cross-order coupling. Magnetoelectric couplings can be 
used, for example, to design ultra efficient logic and memory devices and realize large piezomagnetic 
coefficients for the miniaturization of antennas and sensors. Furthermore, phase transitions involving 
${\cal T}$ phases typically exhibit colossal volume changes of $\sim 10$\% (e.g., PbVO$_{3}$ and related 
solid solutions), which can be exploited in mechanical degradation [\onlinecite{yamamoto19,pan19}] applications. 
Examples of ${\cal T}$ multiferroic materials are bulk BiCoO$_{3}$ (BCO) and BiFeO$_{3}$ (BFO) thin films 
[\onlinecite{menendez20,belik06,wang03}].

Recently, it has been shown by means of first-principles calculations based on DFT that 
BiFe$_{1-x}$Co$_{x}$O$_{3}$ solid solutions (BFCO) with $0.25 \le x \le 0.50$ represent ideal bulk systems
in which to realize the full potential of multiferroic ${\cal T}$ materials [\onlinecite{menendez20b}]. 
Specifically, it has been found that under moderate hydrostatic pressures of $0.1 \lesssim P \lesssim 1$~GPa 
(depending on the relative Fe/Co content) it is possible to trigger a phase transition from a low-$T$ 
rhombohedral (${\cal R}$) phase to a high-$T$ ${\cal T}$ phase (Figs.\ref{fig4}a--c) at room temperature. 
The disclosed pressure-induced ${\cal R} \to {\cal T}$ phase transformation involves (i)~a colossal increase 
in the electric polarization of $\sim 150$\%, (ii)~the existence of a robust net magnetization of $\approx 
0.13$~$\mu_{B}$ per formula unit, and (iii)~a giant volume increase of $\Delta V \sim 10$\% 
(Fig.\ref{fig4}d). Examples of technologies in which these multifunctional phenomena could have an impact 
include pyroelectric energy harvesting [\onlinecite{bowen14,hoffmann15}] and solid-state cooling 
[\onlinecite{gottschall18}].

Figures~\ref{fig4}d and e show respectively the isothermal entropy and adiabatic temperature changes induced 
by hydrostatic pressure in BiFe$_{0.5}$Co$_{0.5}$O$_{3}$ solid solutions. Those results have been obtained with
the quasi-harmonic DFT method (Sec.\ref{subsec:QHA}) and the Clausius-Clapeyron indirect estimation approach 
(Sec.\ref{sec:clausius}) [\onlinecite{menendez22}]. First, it is noted that the ${\cal R} \to {\cal T}$ 
transition temperature can be enormously shifted by small pressure changes (Fig.\ref{fig4}c). For instance, in 
BiFe$_{0.5}$Co$_{0.5}$O$_{3}$ a compression decrease of only $0.06$~GPa moves the transition temperature from 
room temperature down to about $200$~K. Second, an isothermal entropy change of approximately 
$-1.5$~J~K$^{-1}$mol$^{-1}$ (equivalent to $\approx -4.8$~J~K$^{-1}$kg$^{-1}$) is estimated at room temperature 
along with a $\Delta T$ of $\approx 5$~K. These figures, although cannot be regarded as ``giant'', are reasonably large and promising. And third, the size of these two barocaloric descriptors significantly increase at lower 
temperatures. At $T = 200$~K, for instance, $\Delta S$ increases up to $-2.8$~J~K$^{-1}$mol$^{-1}$ and 
$\Delta T$ to $6.6$~K.  

The driving pressures reported in Fig.\ref{fig4}d are very large, of the order of $1$~GPa; however, in practice 
those values can be drastically reduced (i.e., by one order of magnitude, $\sim 0.1$~GPa) by increasing the 
relative content of Fe ions. For instance, by using analogous first-principles computational methods a 
${\cal R} \to {\cal T}$ phase transition pressure of $0.2$~GPa is estimated at zero temperature for bulk 
BiFe$_{0.75}$Co$_{0.25}$O$_{3}$ [\onlinecite{menendez22}]. This theoretical outcome is in good agreement
with the available experimental data on the compositional phase diagram of BFCO solid solutions 
[\onlinecite{hojo18}]. 
Another interesting aspect of BFCO solid solutions is that, besides hydrostatic pressure, they can react to 
external magnetic and electric fields since these are simultaneously magnetic and ferroelectric 
[\onlinecite{menendez20b}]. Such a multifunctional quality converts BFCO solid solutions into a potentially 
ideal playground in which to explore and take advantage of novel multicaloric effects and cycles 
[\onlinecite{cazorla19b,gottschall18}]. Therefore, BFCO solid solutions in particular and multiferroic materials 
exhibiting supertetragonal phases in general, appear to be very promising materials for the development of 
advanced solid-state cooling applications.

\section{Conclusions}
Atomistic computational techniques for the simulation of barocaloric effects are already well 
established. The simulation methods surveyed in this Review (e.g., quasi-harmonic and thermodynamic 
integration techniques) were originally developed for the study of temperature- and pressure-induced phase 
transformations in materials, and have already achieved great success in many research disciplines (e.g., 
high-pressure physics). Nonetheless, the number of atomistic simulation studies found in the literature on
barocaloric effects currently is quite reduced and mostly limited to indirect estimation approaches. 
One of the main objectives of this Review is to improve this situation by surveying key theoretical and 
computational concepts in the estimation of barocaloric effects and to promote the use of alternative
and highly accurate direct and quasi-direct simulation methods.  

As it has been illustrated here, atomistic simulation approaches can be used to reproduce with reliability 
the barocaloric performance of complex materials like multiferroics, in which the structural and electronic 
degrees of freedom are strongly coupled, lithium-based hydrides, in which molecular ions can be orientationally 
disordered, and fast-ion conductors, in which ionic diffusion determines most of their physical properties. 
Analogous barocaloric simulation success can be achieved for other families of functional materials exhibiting 
complex and unconventional, as well as standard, responses to hydrostatic pressure.        

Atomistic simulation of barocaloric effects, therefore, can indeed help enormously in developing new materials 
and strategies for boosting solid-state cooling based on the application of hydrostatic pressure. Thus, 
the current pressing challenge of finding new refrigeration technologies that are environmentally friendly 
and also sustainable may greatly benefit from the reliable and physically insightful computational methods 
surveyed in this Review.

\end{document}